\documentclass[12pt]{article}
\usepackage{amsfonts,latexsym}
\newcommand{\R}{{\mathbb R}}

\let\p\partial

\newtheorem{lem}{Lemma}
\newtheorem{theo}{Theorem}

\begin{document}

\begin{center}
{\large\bf
R.Z. Zhdanov~${}^*$, 
I.M. Tsyfra~${}^\dagger$, 
and R.O. Popovych~${}^\ddagger$
\\[3ex]
{\LARGE \bf A precise definition of reduction\\[2mm] of partial dif\/ferential equations}
\\[5ex]} 
Institute of Mathematics of NAS of Ukraine, \\
3 Tereshchenkivska Street, 01601 Kyiv-4, Ukraine
\end{center}

\noindent
$^{*}$~E-mail: {\tt renat@imath.kiev.ua}, URL: {\tt http://www.imath.kiev.ua/\~{}renat/}\\[2mm]
${}^\dag$~E-mail: {\tt tsyfra@igph.kiev.ua}\\[2mm]
${}^\ddag$~E-mail: {\tt rop@imath.kiev.ua}, URL: {\tt http://www.imath.kiev.ua/\~{}rop/}

\vspace{5ex}

\begin{abstract} \noindent
We give a comprehensive analysis of interrelations between the basic
concepts of the modern theory of symmetry (classical and non-classical)
reductions of partial differential equations. Using the introduced
definition of reduction of differential equations we establish 
equivalence of the non-classical (conditional symmetry) and direct
(Ansatz) approaches to reduction of partial dif\/ferential equations. As
an illustration we give an example of non-classical reduction of the
nonlinear wave equation in $1+3$ dimensions. The conditional symmetry
approach when applied to the equation in question yields a number of
non-Lie reductions which are far-reaching generalization of the
well-known symmetry reductions of the nonlinear wave equations.

\end{abstract}

\section{Introduction}

The notion of a non-classical symmetry was introduced by Bluman and Cole
as early as in 1969 \cite{blu69}. However, non-trivial examples of
non-classical symmetries for nonlinear partial differential equations
(PDEs) appeared much later in the papers by Olver \& Rosenau
\cite{olv86,olv87} and by Fushchych \& Tsyfra \cite{tsy87}. These papers
together with the ones by Fushchych \& Zhdanov, \cite{zhd89a}, Clarkson
\& Kruskal \cite{cla89}, Levi \& Winternitz \cite{win89} gave a start to
an intensive search for non-classical symmetries of a wide range of
nonlinear differential equations. Following the suggestion by Fushchych
\cite{fus87a}--\cite{fus87c} we call this kind of non-Lie symmetries
conditional symmetries (CSs).

The vast majority of the papers devoted to constructing CSs of nonlinear
PDEs consider equations having two independent variables only. This is
explained by the fact that the determining PDEs for CSs are nonlinear
and have the dimension which is equal to the sum of the number of
dependent and independent variables of the PDE under study. That is why,
there is no systematic general procedure for obtaining CSs of
multi-dimensional nonlinear PDEs. Constructing CS for a specific
multi-dimensional differential equation requires preliminary guesswork
enabling one to reduce the dimension of the system of determining PDEs.
In the papers \cite{zhd89a}, \cite{zhd88}--\cite{zhd97a} devoted to
studying CSs of multi-dimensional nonlinear equations of quantum field
theory (wave, Dirac, L\`evi-Leblond and $SU(2)$ Yang-Mills equations) we
developed an efficient approach based on fixing  a special Ansatz for a
conditionally-invariant solution to be found. The underlying idea for
choosing such an Ansatz was a proper use of Lie symmetry properties of
the equation under consideration (a complete account of the results
obtained in this way can also be found in the monograph \cite{zhd97b}).

There exists a number of different approaches to utilizing CSs of PDEs
in order to reduce these to equations with fewer number of independent
variables. However, with all the differences between these methods they
can be classified into two major groups. The first one is composed by
the methods that are close to the traditional Lie approach and can be
regarded as the \lq infinitesimal methods for finding CS\rq\
\cite{blu69}--\cite{zhd89a}, \cite{win89}, \cite{zhd92}--\cite{popov3}.
The central role is played by infinitesimal CSs within the class of
first-order differential operators. Given an operator of conditional
symmetry, we can construct an Ansatz reducing the dimension of PDE under
study. The second group of the methods are the \lq direct\rq\ ones
\cite{cla89}, \cite{zhd88}--\cite{zhd97a}, \cite{fus83a}--\cite{olv94}
(see also \cite{fshs} and the references therein) that goes up,
probably, to the papers by Fourier and Euler devoted to finding
particular solutions of the two-dimensional heat  equation with the help
of a substitution of a special (separated) form. Namely, the methods in
question are based on fixing a special Ansatz for a solution to be
found. As a rule, these Ans\"atze contain arbitrary functions which are
to be so chosen that some reduction requirements must be met. One of the
principal motivations for writing the present article is studying
interrelations between these approaches. A necessary ingredient of such
study is a precise mathematical definition of reduction of PDEs. We
attempt to give this definition which is the core result of the paper.
Based on this definition is our proof of equivalence of the above two
approaches to reduction of PDEs provided some reasonable restrictions
are met (see, also \cite{est95}). The present paper is a natural
continuation of our earlier papers \cite{zhd92,zhts}, where some ideas
presented below were indicated. We present these ideas in a rigorous
mathematical form which, as we believe, should give new insights into
the theory of conditional symmetries of PDEs.

\section{Ans\"atze and involutive sets of operators}

Consider a family of first-order differential operators in the
variables $x\!=\!(x_1,\ldots,x_n),$ $u$
\begin{equation}
\label{2}
Q_a=\sum\limits_{i=1}^n\xi_{ai}(x,u){\p\over\p {x_i}}
+ \eta_a (x, u){\p\over\p {u}},\quad a=1,\ldots,m,
\end{equation}
where $\xi_{ai}, \eta_a$ are some continuously differentiable functions
in an open domain in $\R^{n+1}$, $m<n.$  The variable $u$ is regarded as
dependent, i.e., it corresponds to the function $u=u(x)$. In a sequel, we
suppose that the conditions of the theorem about implicit function are
fulfilled,  wherever applicable.
\vspace{1.5mm}

\noindent
{\bf Definition 1.} Family of first-order differential operators   
(\ref{2}) is called involutive if there exist smooth functions 
$\mu^c_{ab}(x,u),$ $a,b,c=1,\ldots,m,$ such that 
\begin{equation}
\label{3}
[Q_a,\, Q_b]=\sum\limits_{c=1}^m\mu^c_{ab} Q_c, \ \ a,b =1,\ldots,m.
\end{equation}

The simplest example of an involutive family of operators is given by
first-order differential operators forming a Lie algebra. In this case
$\mu^c_{ab}=\mbox{\rm const}$,\ $a,b,c=1,\ldots,m$ are called structure
constants of the Lie algebra. This implication has far-reaching
consequences in the modern theory of non-Lie reductions of PDEs. This is
explained by the fact that involutive families of operators of the form
(\ref{2}) play the same role in the theory of non-classical symmetry
reductions of PDEs having non-trivial conditional symmetries as that
played by finite-dimensional Lie algebras in the theory of symmetry
reductions of invariant PDEs.

In a sequel we consider involutive families of operators (\ref{2})
satisfying an additional constraint
\begin{equation}
\label{4}
{\rm rank}\, \|\xi_{ai}(x,u) \|^{m \ \ \, n}_{a=1\,i=1} =
{\rm rank}\, \|\xi_{ai}(x,u), \eta(x,u) \|^{m \ \  \,n}_{a=1\,i=1}
=m.
\end{equation}

By direct computation we check that, if operators (\ref{2}) form an
involutive family, then the family of differential operators
\begin{equation}
\label{4a}
Q'_a = \sum\limits_{b=1}^m\lambda_{ab}(x,u) Q_b, \quad
{\rm det}\,\|\lambda_{ab}(x,u)\|^m_{a,b=1}\ne 0 
\end{equation}
is also involutive (see, also \cite{zhd92}). Furthermore the involutive 
family (\ref{4a}) is easily seen to obey the condition (\ref{4}).

Provided the relation of the form (\ref{4a}) holds, two involutive
families of operators $\{Q_a\}$ and $\{Q'_a\}$ are called equivalent.
This equivalence relation splits the set of involutive
families of $m$ operators into equivalence classes forming the
quotient set. We denote this set as ${\cal I}$.

It is a common knowledge that conditions (\ref{3}) are 
sufficient for the system of PDEs
\begin{equation}
\label{5}
Y_a(x,u,\mathop{u}\limits_{\scriptscriptstyle 1})=
\sum\limits_{i=1}^n\xi_{ai}(x,u)\frac{\p u}{\p x_i}-\eta_a(x,u)=0, 
\quad a=1,\ldots,m
\end{equation}
to be compatible (the Frobenius theorem \cite{schutz}). Its general
solution can be locally represented in the form
\begin{equation}
\label{6}
F (W_1,\, \ldots, W_{n+1-m}) = 0,
\end{equation}
where $F$ is an arbitrary smooth function of the variables $W_j,$ 
$W_j=W_j(x,u),$ $j=1,\ldots,n+1-m$ are functionally-independent first
integrals of system of PDEs (\ref{5}).

Due to constraint (\ref{4}) there exists a first integral $W_k(x,u)$ such
that the condition $\p W_k/\p u \ne 0$ holds locally, since otherwise
integrals $W_1$,\ $W_2$, $\ldots$, $W_{n+1-m}$ would be
functionally-dependent.

Changing, if necessary, enumeration, we can put $k=1$. Solving
(\ref{6}) with respect to $W_1$ and introducing the notations
\[
\omega(x,u)=W_1(x,u),\quad \omega_j(x,u)=W_{j+1}(x,u),\quad
j=1,\ldots,n-m
\]
we get the following expression:
\begin{equation}
\label{7}
\omega(x,u) = \varphi (\omega_1(x,u),\ldots,\omega_{n-m}(x,u)),
\end{equation}
where $\varphi$ is an arbitrary smooth function of the variables 
$\omega_1,\,\ldots,\,\omega_{n-m}$.
\vspace{1.5mm}

\noindent
{\bf Definition 2.}\ We call an expression of the form (\ref{7}), where
$\varphi$ is an arbitrary smooth function, 
$\omega(x,u),\,\omega_1(x,u),\,\ldots,\,\omega_{n-m}(x,u)$  are
functionally-independent and  $\p\omega/\p u\ne 0$, an Ansatz for the
field $u=u(x)$.

\begin{lem} There is one-to-one correspondence between the
set of Ans\"atze for the field $u=u(x)$ and the elements
of the space ${\cal I}$.
\end{lem}
Proof.\ While constructing the general solution of system (\ref{5})
we have shown that each involutive family obeying (\ref{4}) gives
rise to the Ansatz of the form (\ref{7}). Furthermore, by construction
equivalent involutive families of operators have the same set
of functionally-independent first integrals. Hence we conclude
that each element of ${\cal I}$ corresponds to one and only one Ansatz
(\ref{7}).

Let us prove the inverse, namely, that each Ansatz (\ref{7})
corresponds to one and only one element of the space ${\cal I}$.
Choose the functions $\theta_a(x,u), a=1,\ldots,m$  so that the
expressions
\begin{equation}
\label{8}
\theta_a(x,u),\quad \omega(x,u),\quad \omega_j(x,u),\quad
a=1,\ldots,m,\ j=1,\ldots,n-m
\end{equation}
are functionally-independent. Then the functions (\ref{8})
form the new coordinate system in the space of variables
$x, u$
\begin{eqnarray}
&&y_a=\theta_a(x,u),\quad z_j=\omega_j(x,u),\quad v=\omega(x,u), \label{9}\\
&&a=1,\ldots,m,\quad j=1,\ldots,n-m. \nonumber
\end{eqnarray}

Rewriting (\ref{7}) in the new variables $y, z, v$ we arrive at
the following expression:
\begin{equation}
\label{10}
v=\varphi(z_1,\ldots,z_{n-m}).
\end{equation}

Evidently, the formula (\ref{10}) give the general solution
of the system of PDEs $\p v/\p y_a=0,$ $a=1,\ldots,m$. The operators
$Q_a=\p/\p y_a, a=1,\ldots,m$ form an involutive family
(since they commute) and fulfill the condition (\ref{4}).
These properties are preserved after rewriting the operators
$Q_a$ in the initial variables $x, u$.

Thus we have constructed an involutive family of operators which
corresponds to a given Ansatz for the field $u=u(x)$. However,
this correspondence is not one-to-one, since the choice of
the functions $\theta_a(x,u)$ is ambiguous. Let us show that choosing
another set of functions $\chi_1(x,u),\ldots,\chi_m(x,u)$ will lead
to an involutive family which is equivalent to the above obtained
involutive family.

Indeed, consider the transformation of variables
\begin{eqnarray}
&&y'_i=\chi_a(x,u),\quad z'_j=\omega_j(x,u),\quad
v'=\omega(x,u), \label{9a} \\
&& a=1,\ldots,m,\quad j=1,\ldots,n-m \nonumber
\end{eqnarray}
which reduces the initial Ansatz to become (\ref{10}). Comparing
(\ref{9}) and (\ref{9a}) we conclude that the map relating the
coordinate systems $y, z, v$ and $y', z', v'$ is of the form
\[
y'_a=F_a(y,z,v),\quad z'_j=z_j,\quad v'=v
\]
with $a=1,\ldots,m$, $j=1,\ldots,n-m$. Consequently,
the operators $\p/\p y_a$ after being rewritten
in the new variables $y', z', v'$ read
\[
{\p \over \p y_a}=
\sum\limits_{b=1}^m{\p y'_b \over \p y_a}{\p \over \p y'_b}=
\sum\limits_{b=1}^m\frac{\p F_b}{\p y_a}{\p \over \p y'_b}=
\sum\limits_{b=1}^m\frac{\p F_b}{\p y_a}Q'_b,\quad a=1,\ldots,m.
\]
Since ${\rm det}\, \|\p F_b/\p y_a\|_{a,b=1}^m \ne 0$, hence
it follows that the involutive families $Q_a=\p/\p y_a$
and  $Q'_a=\p/\p y'_a$ are equivalent.

Thus each involutive family that corresponds to a fixed Ansatz for
the field $u=u(x)$ belongs to the same equivalence class. This
is the same as what was to be proved.  $\rhd$
\vspace{2mm}

\section{Conditional symmetry of PDEs}

Consider PDE of the form
\begin{equation}
\label{1}
L(x, u, \mathop{u}\limits_{\scriptscriptstyle 1}, \ldots,
\mathop{u} \limits_{\scriptscriptstyle r}) = 0,
\end{equation}
where $x = (x_1, x_2, \ldots, x_{n}),$  $u = u(x)$ is a sufficiently
smooth function and the symbol $\mathop{u}\limits_{\scriptscriptstyle
s}$ stands for the set of partial derivatives of the function $u(x)$ of
the order $s$, i.e.,
\[
\mathop{u}\limits_{\scriptscriptstyle s}=
\left\{u_{i_1\ldots i_s}\stackrel{\mbox{\scriptsize def}}{=}
{\p^s u\over \p x_{i_1}\ldots \p x_{i_s}},
\ 1\le i_1 \le n,\,\ldots,\, 1\le i_s \le n
\right\}. 
\]

Within the local approach (used throughout the paper) PDE (\ref{1}) is
treated as an algebraic equation  in the jet space $J^{(r)}$ of the
order $r$. Then $L$ is a smooth function from ${\cal D}$ into $\R,$
where ${\cal D}$ is an open domain in $J^{(r)}$. 

Denote the manifold defined by the equation $L=0$ in $J^{(r)}$ by 
${\cal L},$ the set of all differential consequences of the system
of PDEs (\ref{5}) of the order not higher than $r-1$ (we remind
that $r$ is the order of the initial equation (\ref{1})) by the
symbol $M$ and the corresponding manifold in $J^{(r)}$ by ${\cal M}.$

The most widely used definition of conditional invariance
is the following one.
\vspace{2mm}

\noindent
{\bf Definition 3.}\ PDE (\ref{1}) is conditionally-invariant
with respect to involutive family of operators (\ref{2}) if the
relation 
\begin{equation}
\label{11a}
\mathop{Q}\limits_{\scriptscriptstyle (r)}\!{}_a L\:
\mbox{\raisebox{-0.7ex}{$\Bigr|_{\:{\cal L}\cap{\cal M}}$}}=0
\end{equation}
holds $\forall a=1,\ldots,m$. Here the symbol
$\mathop{Q}\limits_{\scriptscriptstyle (r)}$ stands for the $r$th
prolongation of the operator $Q_a$.

This definition is very useful when computing CSs for a specific PDE.
However, for theoretical considerations it is preferable to utilize the
alternative definition of conditional invariance given below.
\vspace{2mm}

\noindent
{\bf Definition 4.}\ PDE (\ref{1}) is conditionally-invariant
with respect to involutive family of operators (\ref{2}) if the
relation 
\begin{equation}
\label{11b}
\mathop{Q}\limits_{\scriptscriptstyle (r)}\!{}_a \Lambda\:
\mbox{\raisebox{-0.7ex}{$\Bigr|_{\:{\cal L}\cap{\cal M}}$}}=0,
\quad \mbox{where} \quad \Lambda=L\:
\mbox{\raisebox{-0.7ex}{$\Bigr|_{\:{\cal M}}$}},
\end{equation}
holds $\forall a=1,\ldots,m$. It will be shown below that Definitions
3 and 4 are equivalent. Note that there are some other
ways to define CS \cite{blu69,olv86,win89,vor86}, however
Definition 4 is the most convenient for the purposes of this paper.
\vspace{2mm}  

\noindent
{\bf Note.}\ Definitions 3, 4 make sense provided ${\cal L}\cap{\cal
M}\not =\emptyset$. If this is not the case, namely, if there exists an
involutive family such that ${\cal L}\cap{\cal M} =\emptyset$, then we
suppose by definition that PDE (\ref{1}) is conditionally invariant with
respect to this family.

\begin{lem}
Let system of PDEs (\ref{1}) be conditionally-invariant under 
involutive family of differential operators (\ref{2}). Then, it is
conditionally-invariant under involutive family (\ref{4a}) with
arbitrary smooth functions $\lambda_{ab}$.
\end{lem}
{\bf Proof.}\ To prove the lemma we use the special
representation for the coefficients of the $s$th prolongation
of a first-order operator $Q$ given in \cite{olver}:

\vspace{-2mm}

\[
\mathop{Q}\limits_{\scriptscriptstyle (s)}=Q+
\sum\limits_{k=1}^s\,\sum\limits_{i_1,\ldots,i_k=1}^n
\eta_{i_1\ldots i_k}\frac{\p}{\p u_{i_1\ldots i_k}}
\qquad \mbox{if} \qquad 
Q=\sum\limits_{i=1}^n\xi_i(x,u)\frac{\p}{\p x_i}
+\eta(x,u)\frac{\p}{\p u},
\]

\vspace{-3mm}

\noindent where

\vspace{-3mm}

\[\!\!\!
\begin{array}{l}
\eta_{i_1\ldots i_k}=D_{i_1}\ldots D_{i_k}
\Bigl(\eta-\sum\limits_{i=1}^n\xi_iu_i\Bigr)+\xi_i u_{i_1\ldots i_k i},
\\[1.5ex]
i_1,\ldots,i_k=1,\ldots,n, \quad k=1,\ldots,s
\end{array}
\]
and
\[
D_i=\frac{\p}{\p x_i}+u_i\frac{\p}{\p u}
+\sum\limits_{p=1}^\infty\sum\limits_{i_1,\ldots,i_p=1}^n
u_{i_1\ldots i_p i}\,\frac{\p}{\p u_{i_1\ldots i_k}},
\quad i=1,\ldots,n
\]

\vspace{-2mm}

\noindent
is a total differentiation operator with respect to the
variable $x_i$.

Using the above identity yields the chain
of equations that correspond to condition (\ref{11a}) for 
the operators (\ref{4a}):
\begin{equation}\label{chain.1}
\mathop{Q}\limits_{\scriptscriptstyle (r)}\!{}'_a L\:
\Biggr|_{\:{\cal L}\cap{\cal M}}
=\:
\Biggl(\sum\limits_{b=1}^m\lambda_{ab}(x,u)
\mathop{Q}\limits_{\scriptscriptstyle (r)}\!{}_b L\Biggr)\:
\Biggr|_{\:{\cal L}\cap{\cal M}}
=\:
\sum\limits_{b=1}^m\lambda_{ab}(x,u)
\Biggl(\mathop{Q}\limits_{\scriptscriptstyle (r)}\!{}_b L\:
\Biggr|_{\:{\cal L}\cap{\cal M}}\Biggr)=\:0.
\end{equation}

Evidently, the same arguments apply if we use Definition 4. The chain of
equations analogous to the above equations (\ref{chain.1}) is obtained,
where one should replace $L$ by $\Lambda.$ The lemma is proved. $\rhd$
\vspace{2mm}

One of the important consequences of the above lemma is that while
studying conditional symmetry of PDEs we can restrict our considerations
to elements of the quotient space ${\cal I}$. This enables choosing the
most simple representative of each equivalence class in the way
described below.

Let (\ref{2}) be an involutive family of differential operators
satisfying condition (\ref{4}). Then it is possible to choose
the functions $\lambda_{ab}(x,u)$ and, if it is necessary, to 
change enumeration of the variables $x_1,\ldots,x_n$ in such a 
way that operators (\ref{4a}) take the form
\[\!\!\!
\begin{array}{l}\displaystyle
Q'_a=\sum\limits_{b=1}^m\lambda_{ab}
\Biggl(\sum\limits_{i=1}^n\xi_{bi}{\p\over\p x_i}
+\eta_b{\p\over\p u}\Biggr)={\p\over\p x_a} 
+ \sum\limits_{j=m+1}^n\xi_{aj}^\prime{\p\over\p x_j}+
\eta'_a{\p\over\p u},
\\[2ex] a=1,\ldots,m.
\end{array}
\]
Since the family of operators $Q_a^\prime,\ a=1,\ldots,m$ is also
involutive, there exist functions $\tilde\mu^c_{ab}(x,u)$ such that
\[
[Q'_a,\, Q'_b]= \sum\limits_{c=1}^m\tilde\mu^c_{ab} Q'_c,
\ \ a,b=1,\ldots,m.
\]

Computing commutators on the left-hand sides of the above equalities
and equating coefficients of the linearly independent differential
operators $\p/\p x_1$,\ $\ldots$, $\p/\p x_n$ we have
$\tilde\mu^c_{ab} = 0,\ a,b,c=1,\ldots,m$. Consequently, operators
$Q_a^\prime$ form a commutative Lie algebra. Hence, we conclude
that there is a local coordinate system (\ref{9}) such that the
operators $Q_a'$ after being rewritten in the variables
$y, z, v$ read
\begin{equation}
\label{14}
Q'_a={\p\over\p y_a},\quad a=1,\ldots,m.
\end{equation}

Consequently, without loss of generality we may consider
commuting families of operators. This fact simplify calculations,
since the latter can always be represented in the form (\ref{14}). 

\begin{lem}\label{lemma.equivalence.definitions.of.cond.symmetry}
Relation (\ref{11b}) holds true if and only if relation (\ref{11a})
holds true. 
\end{lem}
{\bf Proof.}\ It suffices to consider the case  ${\cal L}\cap{\cal
M}\not=\emptyset$. Let us fix an arbitrary point  ${\bf
j}^0=(x^0,u^0,\mathop{u}\limits_{\scriptscriptstyle 1}\!{}^0,  \ldots,
\mathop{u}\limits_{\scriptscriptstyle r}\!{}^0)\!\in\!J^{(r)}\!:$ 
$\:{\bf j}^0\!\in\!{\cal L}\cap{\cal M}.$

As established above we can suppose without loss of generality that the
operators $Q_a$ commute. Choosing an appropriate coordinate
transformation (\ref{9}) in a neighborhood of $(x^0,u^0)$ we reduce them
to become $Q_a=\p/\p y_a$. Now the manifold ${\cal M}$ is determined by
the following set of $N$ algebraic equations in the space of variables
$y, z, \mathop{v}\limits_{\scriptscriptstyle 1}, \ldots, \mathop{v}
\limits_{\scriptscriptstyle r}$:
\[
{\cal M}=\Bigl\{(y, z, \mathop{v}\limits_{\scriptscriptstyle 1}, \ldots, 
\mathop{v} \limits_{\scriptscriptstyle r})\:\biggl|\:
\begin{array}{c}
\forall s=1,\ldots,r, \;
\forall i_1,\ldots,i_s=1,\ldots,n \\
(\exists k=1,\ldots,s:\:i_k\le m)
\end{array}
\!\!\!:\;v_{i_1\ldots i_s}=0
\biggr\},
\]
where the variable $v_{i_1\!\ldots i_s}$  of the jet space $J^{(r)}$ 
correspond to the derivative $\p^sv/(\p t_{i_1}\!\ldots\p t_{i_s})$, 
$t_a=y_a,\: a=1,\ldots,m$ and $t_j=z_{j-m},\: j=m+1,\ldots,n$.

Taking into account the fact that the relation
\[
L({\bf j}_1)=\Lambda({\bf j}_1)
\]
holds for any point ${\bf j}_1\!\in\!{\cal L}\cap{\cal M}$ and
using the definition of the partial derivative yield the equality
\[
\frac{\p L}{\p y_a}(\,{\bf j}^0)=
\frac{\p\Lambda}{\p y_a}(\,{\bf j}^0).
\]
Since ${\bf j}^0$ is an arbitrary point of ${\cal L}\cap{\cal M}$,
the equation 
\begin{equation}
\label{17}
\frac{\p L}{\p y_a}\:
\mbox{\raisebox{-0.5ex}{$\Biggr|_{\:{\cal L}\cap{\cal M}}$}}=\:
\frac{\p\Lambda}{\p y_a}\:
\mbox{\raisebox{-0.5ex}{$\Biggr|_{\:{\cal L}\cap{\cal M}}$}}
\end{equation}
holds.

Now taking into account the fact that an arbitrary order prolongation of
the operator $\p/\p y_a$ is equal to $\p/\p y_a$ we see that the
left-hand side of (\ref{17}) coincides with the left-hand side of
(\ref{11a}) and the right-hand side of (\ref{17}) coincides with the
left-hand side of (\ref{11b}). Hence it follows the validity of the
assertion of the lemma. $\rhd$

\section{Reduction of PDEs}

We say that Ansatz (\ref{7}) reduces PDE (\ref{1}) if the substitution
of formulae (\ref{7}) into (\ref{1}) gives rise to an equation which is
{\em equivalent} to PDE containing "new" independent  $\omega_{1}$,
$\ldots$, $\omega_{n-m}$ and dependent $\varphi$ variables only. To give
a formal definition let us insert Ansatz (\ref{7}) into the initial
equation (\ref{1}). As a result, we get some $p$th ($p\le r$) order PDE
\[
W(x,u,\varphi,\mathop{\varphi}\limits_{\scriptscriptstyle 1}, \ldots,
\mathop{\varphi} \limits_{\scriptscriptstyle p}) = 0,
\]
where the symbol $\mathop{\varphi}\limits_{\scriptscriptstyle k}$ stands
for the set of $k$th order derivatives of the function $\varphi$ with
respect to the variables  $\omega_{1}$, $\ldots$, $\omega_{n-m}$.
Eliminating the variables $x, u$ with the help of formulae (\ref{8})
yields
\[
W'(\theta_1,\ldots,\theta_m,\omega_1,\ldots,\omega_{n-m},
\varphi,\mathop{\varphi}\limits_{\scriptscriptstyle 1}, \ldots,
\mathop{\varphi} \limits_{\scriptscriptstyle p}) = 0.
\]
{\bf Definition 4.}\ Ansatz (\ref{7}) reduces PDE (\ref{1}) if the
relation
\begin{equation}
\label{15}
\!\!\!\begin{array}{l}
W'=H(\theta_1,\ldots,\theta_m,\omega_1,\ldots,\omega_{n-m},
\varphi,\mathop{\varphi}\limits_{\scriptscriptstyle 1}, \ldots,
\mathop{\varphi} \limits_{\scriptscriptstyle p}){}\\[2ex]
\qquad\quad{}\times\widetilde L(\omega_1,\ldots,\omega_{n-m},
\varphi,\mathop{\varphi}\limits_{\scriptscriptstyle 1}, \ldots,
\mathop{\varphi} \limits_{\scriptscriptstyle p})
\end{array}
\end{equation}
holds with some function $H$ that does not vanish in ${\cal D}\cap{\cal
M}$. The equation $\tilde L=0$ is called the reduced differential
equation. \vspace{2mm}  

\noindent
{\bf Remark}\ The reduced differential equation is determined up  to a
non-vanishing multiplier depending on $\omega_1,\ldots,\omega_{n-m},$
$\varphi,\mathop{\varphi}\limits_{\scriptscriptstyle 1}, \ldots,
\mathop{\varphi} \limits_{\scriptscriptstyle p}.$

\vspace{2mm}  

As mentioned in Introduction there exist two different approaches to
reduction of PDEs that are based on their conditional symmetry. The
first one is solving the determining equations (\ref{11a}) in order to
obtain an involutive family of operators $Q_a$ such that the equation
under study is conditionally-invariant with respect to these operators.
According to \cite{zhd92} an Ansatz corresponding to thus obtained
involutive family reduces the PDE under study in the sense of
Definition~4. Alternatively, one can try to construct an Ansatz
(\ref{7}) reducing the PDE under study without solving an intermediate
problem of finding involutive families of operators obeying (\ref{11a}).
The first approach is usually addressed to as the non-classical or
conditional symmetry reduction method. The second one is called the
Ansatz or direct reduction method. Note that within the framework of the
direct reduction method one always supposes an explicit dependence of an
Ansatz on $u$, thus restricting the choice of Ans\"atze to the following
particular form: \begin{equation} \label{16}
u=f\Bigl(x,\varphi(\omega_1(x),\ldots,\omega_{n-m}(x))\Bigr).
\end{equation} This assumption simplify substantially calculation
involved but, on the other hand, it may result in loosing some classes
of Ans\"atze which have implicit dependence on $u$. This is indeed the
case for the relativistic eikonal equation where some invariant
Ans\"atze cannot be represented in the form (\ref{16}) \cite{fshs}.

Now we are going to prove that the conditional symmetry reduction
and Ansatz approaches are equivalent.

\begin{theo}
Let system of PDEs (\ref{1}) be conditionally-invariant under the
involutive family of differential operators (\ref{2}) satisfying
condition (\ref{4}) and let the function $\Lambda=L\:
\mbox{\raisebox{-0.2ex}{$\bigr|_{\:{\cal M}}$}}$ have the maximal rank
on  ${\cal L}\cap{\cal M}$ or be identically equal to 0.  Then, the Ansatz
(\ref{7}) corresponding to (\ref{2}) reduces system of PDEs (\ref{1}).
Inversely, let Ansatz (\ref{7}) reduce PDE (\ref{1}). Then there is
an involutive family of operators (\ref{2}) obeying (\ref{4}) and
corresponding to Ansatz (\ref{7}) such that PDE (\ref{1}) is
conditionally-invariant with respect to this involutive family.
\end{theo}
{\bf Proof.}\ {\em Conditional symmetry $\Rightarrow$ reduction.}\ 
Let PDE (\ref{1}) be
conditionally-invariant with respect to an involutive family
of differential operators (\ref{2}) obeying the relation
(\ref{4}) and let the function $\Lambda$ have the maximal rank $1$ on 
${\cal L}\cap{\cal M}$. If ${\cal L}\cap{\cal M}=\emptyset$, then
we can choose $H=W'$, $\widetilde L=1$ in (\ref{15}), which
means that PDE (\ref{1}) is reduced to the incompatible equation 1=0. 

Suppose now that ${\cal L}\cap{\cal M}\not=\emptyset$. Using the 
arguments analogous to those applied to prove
Lemma \ref{lemma.equivalence.definitions.of.cond.symmetry}, we rewrite
(\ref{11b}) as follows

\[
\frac{\p\Lambda}{\p y_a}\:
\mbox{\raisebox{-0.5ex}{$\Biggr|_{\:{\cal L}\cap{\cal M}}$}}=0.
\]
Making use of the Hadamard lemma, we represent the above relation
in the equivalent form:
\begin{equation}
\label{18}
\Lambda_{y_a}=F_a\Lambda,\quad a=1,\ldots,m.
\end{equation}

Consequently, the function $\Lambda$ is a solution of the
over-determined system of PDEs (\ref{18}), where $F_a$ are smooth
functions of the variables $y, z, v,
\mathop{v}\limits_{\scriptscriptstyle 1}\!{}^{(z)}, \ldots, \mathop{v}
\limits_{\scriptscriptstyle r}\!{}^{(z)}.$  Here  the symbol
$\mathop{v}\limits_{\scriptscriptstyle s}\!{}^{(z)}$ corresponds to the
set of the $s$-order derivatives of the function $v$ with respect to the
variables $z$ only. The necessary and  sufficient compatibility
conditions of (\ref{18}) read
\begin{equation}
\label{comp.cond.after.Hadamar}
\left(\frac{\p F_a}{\p y_b}-\frac{\p F_b}{\p y_a}\right)\Lambda=0, \quad 
a,b=1,\ldots,m.
\end{equation}
As the function $\Lambda$ has the maximal rank on 
${\cal L}\cap{\cal M}=\{M=0,\:\Lambda=0\},$ 
in an arbitrary neighborhood of any point 
${\bf j}\!\in\!{\cal L}\cap{\cal M}$ in ${\cal M}$ there exists a
point ${\bf j}'$ such that $\Lambda({\bf j}')\not=0.$ In view of this
we get from (\ref{comp.cond.after.Hadamar}) the following system of PDEs:
\[
\frac{\p F_a}{\p y_b}=\frac{\p F_b}{\p y_a}, \quad a,b=1,\ldots,m.
\]
Consequently, there is a function $F$ of the variables $y, z, v,
\mathop{v}\limits_{\scriptscriptstyle 1}\!{}^{(z)}, \ldots, \mathop{v}
\limits_{\scriptscriptstyle r}\!{}^{(z)}$  such that $F_a=\p F/\p y_a,\:
\forall a$. Hence we get the general solution of (\ref{18})
\begin{equation}
\label{19}
\Lambda=\exp\{F\}\widetilde L,
\end{equation}
where $\widetilde L$ is an arbitrary function of 
$z,v,\mathop{v}\limits_{\scriptscriptstyle 1}, \ldots, \mathop{v}
\limits_{\scriptscriptstyle r}$. 

Inserting the Ansatz $v=\varphi(z_1,\ldots,z_{n-m})$ invariant under the
family of operators $\p/\p y_1,\ldots,\p/\p y_m$ into $L$ yields
\[
L\:\mbox{\raisebox{-0.7ex}{$\Bigr|_{\:v=\varphi(z_1,\ldots,z_{n-m})}$}}=
\:\Lambda\:
\mbox{\raisebox{-0.7ex}{$\Bigr|_{\:v=\varphi(z_1,\ldots,z_{n-m})}$}}=\:
\exp\{F\}\widetilde L\:
\mbox{\raisebox{-0.7ex}{$\Bigr|_{\:v=\varphi(z_1,\ldots,z_{n-m})}$}}.
\]

For the case $\Lambda\equiv0$ the proof is obvious (for example, 
we can choose $H\equiv1$ and $L\equiv0$).

\vspace{2mm}

\noindent {\em Reduction $\Rightarrow$ conditional symmetry.}\ 
Let the Ansatz (\ref{7})
reduce PDE (\ref{1}). Let us make the change of variables
(\ref{9}) in order to represent (\ref{7}) in the form (\ref{10}).
Then a corresponding involutive family of differential operators
can be chosen as follows, $Q_a=\p/\p y_a$. Since the
function $\varphi$ is arbitrary, to insert the Ansatz (\ref{10})
into PDE (\ref{1}) written in the new variables $y, z, v(y,z)$ is the
same as to consider the intersection of the manifold ${\cal L}$
with the manifold ${\cal M}$ with a subsequent identifying
$\varphi$ with $v$.

By assumption of the theorem a relation of the form
\[
L\:\mbox{\raisebox{-0.7ex}{$\Bigr|_{\:v=\varphi(z_1,\ldots,z_{n-m})}$}}=\:
H(y,z,\varphi,\mathop{\varphi}\limits_{\scriptscriptstyle 1}, \ldots,
\mathop{\varphi} \limits_{\scriptscriptstyle p}) 
\,\widetilde L(z_1,\ldots,z_{n-m},
\varphi,\mathop{\varphi}\limits_{\scriptscriptstyle 1}, \ldots,
\mathop{\varphi} \limits_{\scriptscriptstyle p})
\]
holds with some non-vanishing $H$. Consequently,
\[
\Lambda\:=\:
L\:\mbox{\raisebox{-0.7ex}{$\Bigr|_{\:\cal M}$}}=\:
H(y,z,v,\mathop{v}\limits_{\scriptscriptstyle 1}\!{}^{(z)}, \ldots,
\mathop{v} \limits_{\scriptscriptstyle p}\!{}^{(z)}) 
\,\widetilde L(z_1,\ldots,z_{n-m},
v,\mathop{v}\limits_{\scriptscriptstyle 1}\!{}^{(z)}, \ldots,
\mathop{v} \limits_{\scriptscriptstyle p}\!{}^{(z)}). 
\]

As the $r$th prolongation of the operator
$Q_a=\p/\p y_a$ is equal to
$\p/\p y_a,\; \forall a$, we have
\[
\mathop{Q}\limits_{\scriptscriptstyle (r)}\!{}_a \Lambda\:=\:
\frac{\p}{\p y_a}\Lambda\:=\:
\frac{\p H(y,z,v,\mathop{v}\limits_{\scriptscriptstyle 1}\!{}^{(z)}, \ldots,
\mathop{v} \limits_{\scriptscriptstyle p}\!{}^{(z)})}{\p y_a}
\,\widetilde L(z_1,\ldots,z_{n-m}, v,
\mathop{v}\limits_{\scriptscriptstyle 1}\!{}^{(z)},
\ldots,\mathop{v}\limits_{\scriptscriptstyle p}\!{}^{(z)}).
\]
Next, as the function $H$ does not vanish in ${\cal D}\cap{\cal M},$
the set of solutions of the equation $\Lambda=0$ coincides with
the set of solutions of the equation $\widetilde L=0$. Consequently, the
relation
\[
\Biggl(\mathop{Q}\limits_{\scriptscriptstyle (r)}\!{}_a \Lambda\:
\Biggr|_{\:\Lambda=0}\,\Biggr)\Biggr|_{\:\cal M}=\:
\mathop{Q}\limits_{\scriptscriptstyle (r)}\!{}_a \Lambda\:
\Biggr|_{\:{\cal L}\cap{\cal M}}=\:0,
\]
holds $\forall a=1,\ldots,m$. Hence, we conclude that the initial
PDE (\ref{1}) written in the variables $y, z, v(y,z)$ is
conditionally-invariant with respect to the involutive family
$\p/\p y_a$. Rewriting (\ref{1}) and the
involutive family $\p/\p y_a$ in the initial
variables $x, u$ completes the proof of the theorem. $\rhd$

\section{Application: the nonlinear wave equation}

Now we are going to consider a specific example enlightening the
peculiarities of the Ansatz (direct) and conditional (non-classical)
symmetry approaches to the problem of dimensional reduction of PDEs. As
a basic model we take the nonlinear (1+3)-dimensional wave equation
\begin{equation}
\label{0.1}
\Box u = F(u).
\end{equation}
Here $\Box = \partial^2/\partial x_0^2 - \Delta$ is the d'Alembertian,
$u=u(x)$ is a real-valued function of four real variables
$x_0,x_1,x_2,x_3$ and $F$ is an arbitrary continuous function.

First, we apply the Ansatz approach to reduction of PDE (\ref{0.1}). To
this end we utilize an idea suggested in \cite{zhd89a} and make use of
the Lie symmetry properties of the equation under study for the sake of
elucidating of a possible structure of the Ansatz for the $u(x)$.

As is well-known the maximal in Lie's sense symmetry group admitted
by Eq.(\ref{0.1}) with an arbitrary $F$ is the  ten-parameter Poincar\'e
group $P(1,3)$ having the generators
\begin{equation}
\label{0.2}
P_\mu={\p\over \p x^\mu},\quad J_{\mu\nu}=x_\mu P_\nu -
x_\nu P_\mu,\quad \mu,\nu=0,1,2,3,\ \mu < \nu.
\end{equation}
Hereafter raising and lowering the indices is performed with the help of
the metric tensor of the Minkowski space $g_{\mu\nu}={\rm diag}\, (1,
-1, -1, -1)$ and the summation convention is used. For example,
\[
x^\mu=g_{\mu\nu}x_\nu=\left\{\begin{array}{ll}
x_0,& \mu=0,\\
-x_a,& \mu=a=1,2,3. \end{array}\right.
\]

Symmetry reduction of Eq.(\ref{0.1}) by subgroups of the Poincar\'e
group has been performed in \cite{fus83b,gru84}. An analysis of
thus obtained invariant Ans\"atze for the scalar field $u(x)$ shows that
they have the same structure
\begin{equation}
\label{0.3}
u(x)=\varphi(\omega(x)).
\end{equation}
The form of a real-valued function $\omega(x)$ is determined
by the choice of a specific subgroup of the group $P(1,3)$.

Thus as a first step of our approach we fix the Ansatz for solutions of
Eq.(\ref{0.1}) to be of the form (\ref{0.3}). However, we do not impose
{\em a priori} restrictions on the choice of unknown function
$\omega(x)$. The only requirement to be met is that inserting the
expression (\ref{0.3}) into Eq.(\ref{0.1}) should yield an ordinary
differential equation (ODE) for the function $\varphi(\omega)$. This
requirement gives rise to a compatible over-determined system of
nonlinear partial differential equations for the functions $\omega(x)$.
Any solution of the latter after being inserted into formula (\ref{0.3})
yields an Ansatz for the scalar field $u(x)$ reducing Eq.(\ref{0.1}) to
ODE.

Inserting (\ref{0.3}) into the nonlinear wave equation
(\ref{0.1}) gives
\begin{equation}
\label{0.4}
\left(\frac{\p\omega}{\p x_\mu}\frac{\p\omega}{\p x^\mu}\right)\,
{d^2\varphi\over d\omega^2}
+ \Box \omega\, {d \varphi\over d\omega} = F(\varphi).
\end{equation}

As the above equation has to be equivalent to ODE for the function
$\varphi(\omega)$ under arbitrary $F$, the coefficients of $d^2\varphi/d
\omega^2,\ d\varphi/ d\omega$ have to be some functions of $\omega$.
This requirement yields that there exist real-valued functions
$f_1(\omega), f_2(\omega)$ such that
\begin{equation}
\label{0.5}
\frac{\p\omega}{\p x_\mu}\frac{\p\omega}{\p x^\mu}=f_1(\omega),\quad
\Box \omega = f_2(\omega).
\end{equation}

System of nonlinear PDEs (\ref{0.5}) is the necessary and sufficient
condition for the Ansatz (\ref{0.3}) to reduce the nonlinear wave
equation (\ref{0.1}) to an ordinary differential equation. And what is
more, the equation for the function $\varphi(\omega)$ reads
\begin{equation}
\label{0.6}
f_1(\omega)\,{d^2\varphi\over d\omega^2} +
f_2(\omega)\,{d \varphi\over d\omega}
= F(\varphi).
\end{equation}

Summing up we conclude that any solution of over-determined system of
nonlinear PDEs (\ref{0.5}) gives rise to an Ansatz for the field $u(x)$
reducing Eq.(\ref{0.1}) to ODE of the form (\ref{0.6}). In particular,
any Ansatz corresponding to the Lie symmetry of the nonlinear wave
equation (\ref{0.1}) can be obtained in this way. However, the Lie 
Ans\"atze do not exhaust the set of all possible substitutions of the
form (\ref{0.3}) reducing Eq. (\ref{0.1}) to ODEs. This is explained by
an existence of wide classes of Ans\"atze (\ref{0.3}) that correspond to
conditional symmetry of the nonlinear wave equation and cannot be, in
principle, obtained within the framework of the Lie symmetry approach.

Now we utilize the conditional symmetry approach for obtaining Ansatz
(\ref{0.3}) that reduces PDE in four dimensions (\ref{0.1}) to ODE.
Consider conditional symmetry of the nonlinear wave equation within the
class of first-order differential operators
\begin{equation}
\label{0.7}
Q=\xi_\mu(x) {\p\over \p x_\mu}.
\end{equation}

As we are looking for conditional symmetries that enable reduction
of (\ref{0.1}) to ODE, it is necessary to consider an involutive
family of three differential operators of the form (\ref{0.7}),
namely, $Q_a=\xi_{a\mu}(x) {\p\over \p x_\mu},\quad a=1,2,3$. And what
is more, we require that the restriction (\ref{4}) is
respected, which means that
\[
{\rm rank}\, \|\xi_{a\mu}(x) \|^{3 \ \ \, 3}_{a=1\,\mu=0} = 3
\]
Taking into account the above relation, Lemma 2 and also making use of
the Poincar\'e invariance of the equation under study we can always
transform the operators $Q_1, Q_2, Q_3$ to become
\begin{equation}
\label{0.8}
Q_a={\p\over \p x_a} - f_a(x){\p\over \p x_0},\quad a=1,2,3.
\end{equation}
It is straightforward to check that the family of operators (\ref{0.8})
is involutive if and only if $Q_1, Q_2, Q_3$ commute each with another.
Hence we get the system of three PDEs for the functions $f_1(x), f_2(x),
f_3(x)$
\[
{\p f_a\over \p x_b} - f_b{\p f_a\over \p x_0}=
{\p f_b\over \p x_a} - f_a{\p f_b\over \p x_0},
\]
where $a,b=1,2,3,\ a<b$. Its general solution can be represented
in the form (see, e.g.,  \cite{popov3})
\begin{equation}\label{representation.for.fa}
f_a={\p \omega\over\p x_a}
\left({\p \omega\over \p x_0}\right)^{-1},\quad a=1,2,3,
\end{equation}
where $\omega=\omega(x)$ is an arbitrary twice continuously
differentiable function,\ $\p\omega/\p x_0\not=0$. 

The condition (\ref{11a}) of invariance of the nonlinear wave equation
(\ref{0.1}) with respect to operators (\ref{0.8}) after some involved
straightforward algebraic manipulations reduces to over-determined
system of six PDEs for the functions $f_1, f_2, f_3$
\begin{equation}
\label{0.9.1}
\Box f_a-2{\p f_a\over \p x_b}{\p f_b\over \p x_0}=0,
\quad
{\p f_a\over \p x_0}-f_b{\p f_a\over \p x_b}=0,\quad a=1,2,3.
\end{equation}
Inserting the expressions for $f_a$ (\ref{representation.for.fa})
into (\ref{0.9.1}) and rearranging the obtained PDEs for the
function $\omega(x)$ yields
\[
\left({\partial \omega\over \partial x_0}{\partial\over\partial x_a}-
{\partial \omega\over \partial x_a}{\partial\over\partial x_0}\right)\,
\frac{\p\omega}{\p x_\mu}\frac{\p\omega}{\p x^\mu}=0,\quad
\left({\partial \omega\over \partial x_0}{\partial\over\partial x_a}-
{\partial \omega\over \partial x_a}{\partial\over\partial x_0}\right)\,
\Box \omega=0
\]
with $a=1,2,3$. Hence we conclude that there are smooth functions
$f_1(\omega), f_2(\omega)$ such that the relations\ $\frac{\p\omega}{\p
x_\mu}\frac{\p\omega}{\p x^\mu}=f_1(\omega),\ \Box
\omega=f_2(\omega)$\ hold and we arrive at system of PDEs (\ref{0.5}).
Consequently, the involutive family (\ref{0.8}) takes necessarily the
form
\begin{equation}
\label{0.9}
Q_a={\p\over \p x_a} - {\p \omega\over\p x_a}
\left({\p \omega\over \p x_0}\right)^{-1}
{\p\over \p x_0},\quad a=1,2,3,
\end{equation}
where the function $\omega=\omega(x)$ is a solution of system (\ref{0.5}).

As the function $\omega=\omega(x)$ is the first integral of the system
of PDEs $Q_af(x)=0,\ a=1,2,3$, the Ansatz for the field $u(x)$
corresponding to the family $Q_1, Q_2, Q_3$ is given by (\ref{0.3}).

Thus both Ansatz (direct) and conditional symmetry (non-classical)
approaches to reduction of the nonlinear wave equation (\ref{0.1}) to
ODEs lead to the same reduction conditions, namely, to the system of
differential equations (\ref{0.1}) consisting of the nonlinear wave and
relativistic Hamilton-Jacobi equations. Following \cite{zhd89b} we call this
system the d'Alembert-Hamilton system.

The d'Alembert-Hamilton system in three dimensions was studied by Jacobi
\cite{bat22}, Smirnov \& Sobolev \cite{smi32,smi33} and later on by
Collins \cite{col76}. Collins constructed the general solution of system
of nonlinear PDEs (\ref{0.4}) for a complex-valued function of three
complex variables. Some exact solutions of the d'Alembert-Hamilton
system in four dimensions have been constructed by Cartan \cite{car55},
Bateman \cite{bat55} and Erugin \cite{eru44}. Recently, we have
constructed the general solution of system (\ref{0.4}) for the
complex-valued function of four complex variables
\cite{zhd90b,zhd91}.

As established in \cite{zhd97b}, system of PDEs (\ref{0.4}) for the
real-valued function $\omega(x)$ is compatible if and only if it is
locally equivalent to the system \begin{equation} \label{0.0} \Box
\omega = \epsilon N\omega^{-1}, \quad
 (\p_\mu \omega) (\p^\mu \omega) = \epsilon,\quad \epsilon=\pm 1,0,
\end{equation}
where $N=0,1,2,3$.

The real form of the general solution of the system of PDEs (\ref{0.0})
is given by one of the formulae below \cite{zhd91b,zhd97b}
\vspace{3mm}

\noindent
{\rm I.}$\quad \epsilon = -1$
\vspace{1.5mm}

\noindent
1)$\quad N = 0$ 
\begin{equation}
\label{n0}
\omega = A_\mu(\tau) x^\mu + R_1 (\tau),
\end{equation}
where $\tau = \tau(x)$ is determined in implicit way
\begin{displaymath}
B_\mu (\tau) x^\mu + R_2(\tau) = 0
\end{displaymath}
and $A_\mu(\tau),\ B_\mu(\tau),\ R_1(\tau),\ R_2(\tau)$ are
arbitrary smooth real-valued functions satisfying the conditions
\begin{displaymath}
A_\mu(\tau) A^\mu(\tau) = -1,\ \ A_\mu(\tau) B^\mu(\tau) = 0,
\ \ \dot A_\mu(\tau) B^\mu(\tau) = 0,\ \
B_\mu(\tau) B^\mu(\tau) = 0;
\end{displaymath}
2)$\quad N = 1$
\begin{eqnarray}
\omega^2 &=&  (d_{\mu} x^{\mu} + g_2)^2 - (a_{\mu} x^{\mu} + g_1)^2,
\label{n1a}\\
\omega^2 &=& (b_\mu x^\mu + C_1)^2 + (c_\mu x^\mu + C_2)^2,\label{n1b}
\end{eqnarray}
where $g_i = g_i(a_\mu x^\mu + d_\mu x^{\mu}) \in C^2({{\bf R}}^1,
{{\bf R}}^1) $ are arbitrary functions;
\vspace{1.5mm}

\noindent
3)$\quad N = 2$
\begin{enumerate}
\item [{a)}]{
\begin{equation}
\label{n2a}
\omega^2 = -\Bigl(x_{\mu} + A_{\mu} (\tau)\Bigr)\Bigl(x^{\mu} +A^{\mu}
(\tau)\Bigr) - \Bigl\{B_{\mu} (\tau)\Bigl(x^{\mu} + A^{\mu}
(\tau)\Bigr)\Bigr\}^2,
\end{equation}
where $\tau = \tau(x) $ is determined in implicit way
\begin{displaymath}
\Bigl(x_{\mu} +A_{\mu} (\tau)\Bigr)\dot B^{\mu}(\tau) = 0,
\end{displaymath}
$A_{\mu} (\tau),\ B_{\mu} (\tau)$ are arbitrary smooth
real-valued functions satisfying the conditions
\begin{displaymath}
B_{\mu}(\tau) B^{\mu}(\tau) = -1,\quad
\dot {B}_{\mu}(\tau) \dot {B}^{\mu}(\tau) =0,
\quad
\dot A_{\mu}(\tau) = R(\tau) \dot B_{\mu}(\tau)
\end{displaymath}
with an arbitrary $R(\tau) \in C^1 ({{\bf R}}^1,\ {{\bf R}}^1)$;}
\item[{b)}]{
\begin{equation}
\label{n2b}
\omega^2 = -\Bigl(x_\mu+A_\mu(\tau)\Bigl)
\Bigr(x^\mu+A^\mu(\tau)\Bigr)
- \Bigl\{b_\mu \Bigl(x^\mu + A^\mu(\tau)\Bigr)\Bigr\}^2,
\end{equation}
where $\tau = \tau(x)$ is determined in implicit way
\begin{displaymath}
\Bigl(x_\mu + A_\mu(\tau)\Bigr)\Bigl(\dot A^\mu (\tau) +
b^\mu b_\nu \dot A^\nu (\tau)\Bigr)= 0,
\end{displaymath}
$A_\mu (\tau)$ are arbitrary smooth real-valued functions
satisfying the condition
\begin{displaymath}
\dot A_\mu(\tau) \dot A^\mu(\tau) +
\Bigl(b_\mu \dot A^\mu(\tau)\Bigr)^2 = 0;
\end{displaymath}}
\item[{c)}]{
\begin{equation}
\label{n2c}
\omega^2 = (b_\mu x^\mu + C_1)^2 + (c_\mu x^\mu + C_2)^2
+ (d_\mu x^\mu + C_3)^2;
\end{equation}}
\end{enumerate}
\vspace{1.5mm}

\noindent
4)$\quad N = 3$
\begin{equation}
\label{n3}
\omega^2 = -\Bigl(x_\mu + A_\mu (\tau)\Bigr)\Bigl(x^\mu + A^\mu
(\tau)\Bigr), 
\end{equation}
where $\tau = \tau(x)$ is determined in implicit way
\begin{displaymath}
\Bigl(x_\mu + A_\mu (\tau)\Bigr)B^\mu (\tau) = 0,
\end{displaymath}
$A_\mu (\tau),\ B_\mu(\tau) $ are arbitrary smooth real-valued
functions satisfying the conditions 
\begin{equation}
\label{n3a}
\dot A_\mu(\tau) B^\mu(\tau) = 0, \quad
B_\mu(\tau) B^\mu(\tau) = 0.
\end{equation}
\vspace{3mm}

\noindent
{\rm II.}$\quad \epsilon =1$
\vspace{1.5mm}

\noindent
1)$\quad N = 0$
\begin{equation}
\label{p0}
\omega = a_\mu x^\mu + C_1;
\end{equation}
\vspace{1.5mm}

\noindent
2)$\quad N = 1$
\begin{equation}
\label{p1}
\omega^2 = (a_\mu x^\mu + C_1)^2 - (d_\mu x^\mu + C_2)^2;
\end{equation}
\vspace{1.5mm}

\noindent
3)$\quad N = 2$
\begin{equation}
\label{p2}
\omega^2 = (a_\mu x^\mu + C_1)^2 - (c_\mu x^\mu + C_2)^2 -
      (d_\mu x^\mu + C_3)^2;
\end{equation}
\vspace{1.5mm}

\noindent
4)$\quad N = 3$
\begin{equation}
\label{p3}
\omega^2 = (x_\mu + C_\mu)(x^\mu + C^\mu).
\end{equation}

\noindent
{\rm III.}$\quad \epsilon = 0,$ $ N = 0$
\[
A_\mu(\omega)x^\mu + B(\omega)=0,
\]
where $A_\mu, B$ are arbitrary smooth real-valued functions such that
$A_\mu A^\mu=0$.

In the above formulae $C_0,\ldots,C_3$ are arbitrary real constants
and $a_\mu$, $b_\mu$, $c_\mu$, $d_\mu$ are arbitrary real constants
satisfying the conditions
\begin{eqnarray*}
&&a_\mu a^\mu = - b_\mu b^\mu = - c_\mu c^\mu = - d_\mu d^\mu = 1,\\
&&a_\mu b^\mu = a_\mu c^\mu = a_\mu d^\mu = b_\mu c^\mu =
b_\mu d^\mu = c_\mu d^\mu = 0.
\end{eqnarray*}

Let us emphasize that all the functions $\omega(x)$ defined by formulae
(\ref{n0}), (\ref{n1a}), (\ref{n2a}), (\ref{n2b}), (\ref{n3}) give rise
to conditionally-invariant Ans\"atze for the field $u(x)$ of the form
(\ref{0.3}). Using these one can construct broad families of new
(non-Lie) exact solutions even for such a well studied model as the
nonlinear wave equation (see, also \cite{zhd98}). Consider, for example,
the confomally-invariant nonlinear wave equation
\begin{equation}
\label{cub}
\Box u= \lambda u^3.
\end{equation}
The Ansatz
\[
u=\varphi\left(-\Bigl(x_\mu + A_\mu (\tau)\Bigr)\Bigl(x^\mu + A^\mu
(\tau)\Bigr)^{1/2}\right),
\]
where $\tau=\tau(x)$ is defined in (\ref{n3}) and $A_\mu(\tau),
B_\mu(\tau)$ are arbitrary smooth functions satisfying (\ref{n3a}),
reduces (\ref{cub}) to ODE for $\varphi=\varphi(\omega)$
\[
{d^2\varphi\over d\omega^2} + 3\omega^{-1}\, {d \varphi\over d\omega}
= -\lambda \varphi^3.
\]
Two particular solutions of the latter $\varphi=\lambda^{-1/2}\omega^{-1}$
and $\varphi=a(\omega^2+\lambda a/8)^{-1}$, $a=$const give rise to two
families of new exact solutions of the cubic wave equation (\ref{cub})
\begin{eqnarray*}
u(x)&=&\lambda^{-1/2}\left[-\Bigl(x_\mu +
A_\mu (\tau)\Bigr)\Bigl(x^\mu + A^\mu (\tau)\Bigr)\right]^{-1/2},\\
u(x)&=&a\left[\frac{\lambda a^2}{8} - \Bigl(x_\mu + A_\mu (\tau)\Bigr)
\Bigl(x^\mu + A^\mu (\tau)\Bigr)\right]^{-1}.
\end{eqnarray*}
Choosing arbitrary functions $A_\mu(\tau)$ to be constant yields the
well-known exact solutions of (\ref{cub}) obtained in \cite{gru84}
within the symmetry reduction routine. However, if $A_\mu(\tau)$ are
not constants, the constructed solutions are new and cannot be found
using the symmetry reduction procedure.

\section{Concluding Remarks}

Thus introducing a rigorous definition of reduction of PDEs enables a
systematic treatment of the problem of studying interrelations between
the Ansatz (direct) and non-classical (conditional symmetry) approaches
to dimensional reductions of multi-dimensional PDEs. We have proved that
the direct approach, taken in a full generality, is equivalent to the
non-classical approach provided some natural restrictions are met (see
Theorem 1). When we say \lq in a full generality\rq\ we mean that the
most general form of the similarity Ansatz should be taken. For example,
the Ansatz (\ref{0.3}) is a particular case of the general similarity
Ansatz for PDE
(\ref{0.1})
\begin{equation}
\label{implicit}
U(x,u)=\varphi(\omega(x,u)).
\end{equation}
Imposing the restrictions $U(x,u)=u, \omega(x,u)=\omega(x)$ results in
loosing some reductions. On the other hand, with this choice of the form
of the Ansatz we were able to get a full solution of the problem of
constructing the corresponding conditionally-invariant Ans\"atze of the
form (\ref{0.3}) (see, formulae (\ref{n0})--(\ref{p3})), as the system
of nonlinear determining equations for $\omega(x)$ proves to be
integrable. Integrating it yields broad classes of principally new
reductions and exact solutions for nonlinear wave equations containing
several arbitrary functions of one argument.

So both direct and nonclassical approaches can be used on equal footing
and the choice of one of them is, in fact, a matter of taste.
Nevertheless, the direct approach has an evident benefit of being
comparatively simple, since only some basics of the standard university
course on partial differential equations are required for understanding
and implementing it. Another merit of the direct approach is its
flexibility. A similarity Ansatz can be easily modified in order to
yield, for example, \lq nonlinear separation of variables\rq\ in the
spirit of \cite{galak} (see, also \cite{zhd95,nucci}). However, if we
wish to take into consideration the case of implicit Ans\"atze (say, of
the form (\ref{implicit})), then the nonclassical approach is
preferable. These points are illustrated by the considerations of
Section 5, where both approaches are applied to the nonlinear wave
equation and the direct method provides a shorter way to obtain
conditional symmetries of the equation under study. 

The fact that we restrict our considerations to scalar PDEs, namely, to
PDEs with one dependent variable, is explained by the major difficulties
arising when handling systems of PDEs. The first problem is the fact
that different equations of system may have different orders. Next, if
the number of equations is greater than the number of dependent
variables, there arises a natural question of compatibility of this
system. However, the implication {\em conditional invariance}
$\Rightarrow$ {\em reduction} can be proved in almost the same way as it
is done for the case of a single PDE in Theorem 1 \cite{zhd92}. The
problem is how to modify the proof in order to establish a validity of
an assertion {\em reduction} $\Rightarrow$ {\em conditional invariance}.
We postpone the investigation of this problem to our future
publications.

\section*{Acknowledgments} This work is partially supported by the
Ukrainian DFFD Foundation under the project 1.4/356.

\end{document}